\documentclass[twocolumn,showpacs,preprintnumbers,amsmath,amssymb]{revtex4}
\usepackage{graphicx}
\usepackage{dcolumn}
\usepackage{bm}
\usepackage{amsfonts}

\begin{document}
\makeatletter

\makeatother

\title{Decoherence of Anyon Qubit}
\author{A. Dementsov}
\address{Center for Quantum Device Technology, Department of
Physics, Clarkson University, Potsdam, NY 13699-5721, USA, e-mail:
dementav@clarkson.edu}

\date{\today}

\begin{abstract}

One mechanism of decoherence of anyon qubit due to interaction
with edge states is considered. The calculations are made at low
temperature in Markovian and "short-time" approximation. Two
approximations are compared.
\end{abstract}

\pacs{03.65.Yz, 03.67.-a}

\maketitle

\section{Introduction}

Quantum computation today is a fast and extensively developing
field of investigations both theoretical and experimental. The
main problem which stands on the way to implementation of
effective qubit for a quantum computer is {\em decoherence} caused
by interaction of the microscopic system with the rest of the
world. It means that degrees of freedom of the qubit system in the
process of its evolution {\em entangle} inevitably with huge
number of other degrees of freedom leading to loss of information
transferred to the system before. On the language of operators,
 density matrix of the world evolve by unitary time evolution operator, where
as density matrix of microscopic system under consideration almost
always does not \cite{DiVincenzo}. Decoherence of a quantum system
has a fundamental nature and actually the main reason of
transition from quantum to classical mechanics. In spite of the
chosen quantum system, this effect shall be taken into
consideration. The question arises here: "What system to choose or
to find so that small enough effect of decoherence could allow
quantum computations"?

One of the possible answers to above question can be found in
quantum systems intrinsically fault-tolerant in the sense that
they are stable to external influence. One of these qubit systems
based on anyons (anyon qubit) was first proposed by Kitaev in
\cite{Kitaev} and then developed in \cite{Lloyd} and
\cite{Freedman}. Possible experimental implementation was
considered by Averin and Goldman in \cite{Averin} for the system
consisting of two antidots in Quantum Hall regime at filling
factor $\nu=1/3$. In spite of challenging idea of the experimental
implementation of anyon qubit there arise many problems standing
on the way of use of the qubit for quantum computations. One of
the main problem, I suppose, is possible measurement of such
system.

Many experiments had been carried out on tunnelling of anyons
trough one antidot in FQH regime \cite{Maasilta-1} -
\cite{Sachrajda}. It was proved \cite{Goldman} that quasiparticles
participating in tunnelling have charge $e/3$ for $\nu=1/3$  which
gives some evidence in anyon charact er of the transport. But I am
not familiar with experiments which explore transport of anyons on
two anidots.

In this article only one mechanism of decoherence of anyon qubit
due to interaction with edge states is presented. The calculations
are made at low temperature  in Markovian and "short-time"
approximation.

\section{Description of the system anyon qubit and
edge states}

In case of weak interaction the total hamiltonian of the system
anyon qubit and edge states can be described in standard way:
\begin{equation}\label{1}
    H = H_{S}+H_{B}+H_{int}
\end{equation}
where $H_{S}, H_{B}$ and $H_{int}$ are hamiltonians of
correspondingly the anyon qubit (Sistem), the edge states (Bath)
and interaction between them.

Anyon qubit consists of two antidots separated by distance $d$.
Applying voltage to the gates near one or the other antidot it is
possible to control location of anyon on one of the two antidots.

Tunnelling of anyons from one antidon to the other is similar to
that of tunnelling of an electron from one well to the other one,
which has been in details reviewed in \cite{Leggett}. By analogy
the Hamiltonian for the model of anyon localization on two
antidots $H_S$ can written as:

\begin{equation}\label{2}
    H_{S}= \frac{1}{2}\varepsilon\sigma_z - \frac{1}{2}\Omega\sigma_x,
\end{equation}

The first term in (\ref{2})describes localization of the anyon
either on the one or the other antidot with energy difference
$\varepsilon$. The second term describes tunnelling of the anyon
between two antidots with energy splitting $\Omega$. We choose
units where $\hbar=1$ and then return to ordinary units in final
formulas.

The second term in (\ref{2}) describes bath modes or edge states.
In the literature these states are also known as Luttinger liquid
or 1D chiral modes. In our case we will use hydrodynamic model
proposed by Wen in \cite{Wen}. According to this model, 2DEG is
considered to be as noncompressible liquid with 2D electron
density $n_e =$ const. Disturbance of this liquid causes emergence
of edge excitations, or waves, described by $\rho(x)=n_eh(x)$ -
linear density along the edge ($h(x)$ - displacement of the
excitation perpendicular to the edge).

It is easy to show that hamiltonian for such excitations in
Fourier representation can be written as \cite{Wen}:

\begin{equation}\label{3}
    H_{B}=\frac{2\pi v}{\nu}\sum_{k>0}\rho_k\rho_{-k}
\end{equation}
The final term $H_{int}$ in (\ref{1}) is interaction between anyon
qubit and edge states. It can be found as work produced by
electric field of anyon qubit on the transfer of electron along
the edge:

\begin{equation}\label{4}
    H_{int}= \sigma_{z}\int dx U(x)\rho(x)
\end{equation}
$U(x)$ - potential difference due to localization of anyon on one
or the other antidot.  For the simplicity reasons we consider edge
excitation moving along the straight line perpendicular to anyon
qubit and set the length of the edge as a unit. As it is easy to
check the length is eliminated from final results.

Quantization of $\rho(x)$ can be made based on the following
commutation relations:

\begin{equation}\label{5}
\begin{split}
    &[\rho_k,\rho_{k'}]=\frac{\nu k}{2\pi}\delta_{k,-k}\vspace{5pt}\\
    &k, k'=\text{integer}\times 2\pi\vspace{5pt}\\
    &[H_B,\rho_k]=vk\rho_k \vspace{5pt}\\
\end{split}
\end{equation}
Based on the last commutation relation we can see that $\rho_k$
and $\rho_{-k}$ behave as boson operators of creation and
destruction respectively:

\begin{equation}\label{6}
    \rho_k = \sqrt{\frac{\nu k}{2\pi}}b^{+}_{k} \hspace{10pt} \rho_{-k} =
    \sqrt{\frac{\nu k}{2\pi}}b_{k}\hspace{10pt} k > 0
\end{equation}
The above hamiltonian now can be rewritten as

\begin{equation}\label{7}
    H = \frac{1}{2}\varepsilon\sigma_z - \frac{1}{2}\Omega\sigma_x +
    \sigma_z\sum_{k>0}\beta_k(b^+_k + b_k) + \sum_{k>0}\varepsilon_k b^+_kb_k
\end{equation}

\begin{equation}\label{8}
    \beta_k = \sqrt{\frac{\nu k}{2\pi}}U_k, \hspace{0.5cm} \varepsilon_k
    = vk,
\end{equation}
where $U_k$ is Fourier transformation of $U(x)$ and $\nu$ -
filling factor of Landau level. As it is easy to see edge states
have linear dispersion law, similar to that of the acoustic
phonons in lattice. It means that edge states are gapless
excitations.

\section{Decoherence of the anyon qubit in Markovian
approximation}

Now let's proceed to calculations of decoherence and consider the
case when $\varepsilon = 0$. Master equation for density matrix
$\rho = \rho_S\otimes\rho_B$ of the system "anyon qubit" and "edge
states" is

\begin{equation}\label{9}
    \dot{\rho}=-i[H,\rho]
\end{equation}
Tracing over bath modes ($\rho_S=Tr_B\rho$) we can get equation
for anyon qubit density matrix (see e.g. \cite{Kapmen}):

\begin{equation}\label{10}
    \begin{split}
    &\dot{\rho}_S = -i[H_S,\rho_S] - \int_0^{\infty}d\tau
   \{\langle BB(-\tau)\rangle \cdot\\
   & \cdot[S, S(-\tau)\rho_S] - \langle B(-\tau)B\rangle [S, \rho_SS(-\tau)]\}\\
    \end{split}
\end{equation}
Here $\langle\ldots\rangle=Tr_B(\ldots\rho_B)$, $[O_1,O_2]$ -
commutator between operators $O_1$ and $O_2$
\begin{equation}\label{11}
    \begin{split}
    &H_S=- \frac{1}{2}\Omega\sigma_x\\
    &S=\sigma_z,\\
    &B=\sum_{k>0}\beta_k(b^{+}_k + b_k)\\
    &O(t)=e^{it H_S}Oe^{-it H_S}\\
    \end{split}
\end{equation}

The Markovian approximation is also known as long-time
approximation and the following assumptions are to be satisfied
for the density matrix:

\begin{enumerate}
    \item $\rho(t^{\prime}) = \rho(t)$ - It means that the system loses
all memory of its past
    \item $\bar{\rho}(t) = \rho_{S}(t)\rho_B(0),$ where $\rho_B(0) = \exp(-\beta
    H_B)/Z$ that is obeys the Gibbson distribution. It means that energy
    by the system never returns again to the system and any
    changes in it doesn't effect on the "bath"
    \item $\langle BB(-\tau)\rangle \rightarrow 0$ when $t \gg
    \tau_0$, where $\tau_0$ is often called {\em correlation time} for the bath. This assumption allows us to replace integral limits
    to infinity
\end{enumerate}

After simple enough calculations the master equation for
$\rho_S=\frac{1}{2}(1+x\sigma_x+y\sigma_y+z\sigma_z)$ can be
rewritten as

\begin{equation}\label{12}
    \begin{split}
    \dot{x} &= - \Gamma x + \lambda\\
    \dot{y} &= (\Omega + \omega)z - \Gamma y\\
    \dot{z} &= -\Omega y\\
    \end{split}
\end{equation}
Here
\begin{equation}\label{13}
\begin{split}\vspace{20pt}
&\Gamma =
\frac{1}{v}\beta^2_{k=\Omega/v}\coth\left(\frac{\Omega}{2T}\right)\\
&\lambda = \frac{1}{v}\beta^2_{k=\Omega/v}\\
&\omega=\frac{2\Omega}{\pi}
\int_{0}^{\infty}dk\beta^2_k\coth\left(\frac{\varepsilon_k}{2T}\right)\frac{1}{\Omega^2-\varepsilon^2_k},\\
\end{split}
\end{equation}
where $\Gamma$ -  dissipation rate. Solving the system of
differential equations it is easy to see that

\begin{equation}\label{13-1}
\begin{split}
    Tr_S(\rho^2_S(t))&=\frac{1}{2}[1+x^2+y^2+z^2]\\
    &=\frac{1}{2}[1+\tanh^2\left(\frac{\Omega}{2T}\right)+C(T)e^{-\Gamma
    t}],\\
\end{split}
\end{equation}
where $C(T)$ - constant magnitude which depends only on initial
conditions and temperature. The constant satisfies the following
conditions: $|C(T)|<1$ and $C(T)\rightarrow 0$ when $T\rightarrow
0$.

Specifically for the case of unscreened field of anyon qubit with
$\nu=1/m$ and $T=0$
\[
U(x)=\frac{q^2d}{2\pi\epsilon\epsilon_0}\frac{1}{r^2}.
\]
$q$ is the charge of anyon qubit. According to the theory of
anyons (see e.g. \cite{Wilczek}) in FQHE $q=e\nu=e/m$. Then
dissipation rate $\Gamma$ in conventional units will be equal to

\begin{equation}\label{14}
    \Gamma=\left(\frac{d}{L}\right)^2\left(\frac{e^2}{2\epsilon\epsilon_0\hbar v}\right)^2\frac{\Omega}{2\pi m^3\hbar
    }e^{-2\Omega L/\hbar v}
\end{equation}
Here $L$ is the distance between the qubit and the edge, $d$ -
distance between two antidots and $\varepsilon$ - dielectric
constant. This formula is different from that of obtained in
\cite{Averin}. For experimental values $\epsilon\simeq 10, v\simeq
10^5$ m/s, $\Omega\simeq 0.1$K, $d\simeq 100$nm and $L\simeq
3\mu$m we have $\hbar\Gamma/\Omega\simeq 10^{-3}$.

\section{Decoherence of the qubit in a short-time
approximation}

For the short-time approximation let's use the formula given in
\cite{Privman}.

\begin{equation}\label{15}
    Tr_S(\rho^2_S(t))=\frac{1}{2}[1+e^{-2B^2(t)}]
\end{equation}

\begin{equation}\label{16}
    B^2(t)=8\sum_{k>0}\frac{\beta^2_k}{\varepsilon^2_k}\sin\left(\frac{\varepsilon_kt}{2}\right)\coth\left(\frac{\varepsilon_k}{2T}\right) =
    AI(t),
\end{equation}
where
\begin{equation}\label{17}
    A=\frac{2}{m^3}\left(\frac{d}{L}\frac{e^2}{2\pi\epsilon\epsilon_0\hbar v}\right)^2
\end{equation}

For this particular system of anyon qubit we have case of Ohmic
dissipation, for

\begin{equation}\label{18}
    \beta^2(x)\sim xe^{-x/\omega_c}
\end{equation}

We are interested in the case when temperature of the system $T$
is lower than any energy scales in the system including
characteristic cut-off frequency $\omega_c=v/4L$ that is $T\ll
\omega_c$. Integral $I(t)$ has the form
\begin{equation}\label{19}
    I(t)=\int_{0}^{\infty}\frac{dx}{x}e^{-x/\omega_{c}}\sin^2(xt)\coth\frac{x}{T}
\end{equation}
and doesn't have analytical solution. But it can be estimated in
asymptotic approximations.

\noindent 1. The case $t\ll 1/\omega_c$ corresponds to
short-times, when characteristic times of system evolution are
much lower than inverse frequencies of the bath modes.
\[
I(t)\simeq\omega_c^2t^2
\]

\noindent 2. The case $ 1/\omega_c\ll t\ll 1/T$ corresponds to
intermediate times. The integral can be approximated as
\[
I(t)\simeq\frac{1}{2}\text{Ei}(1/\omega_ct)\simeq
\frac{1}{2}\ln\omega_ct,
\]
where Ei$(x)$ - exponential integral function.

\noindent 3. The case $1/T \ll t$ corresponds to long times much
more than any inverse frequencies.
\[
I(t)\simeq\pi Tt
\]

We see here that $B^2(t)$ increases quadratically for short times,
logarithmically for intermediate times and linearly for long
times.

Summarizing the asymptotics obtained above the equation (\ref{15})
can be rewritten as

\begin{equation}\label{20}
\begin{split}
    &Tr_S(\rho^2_S(t))\\
    &\simeq\left\{
\begin{array}{ll}
    \dfrac{1}{2}[1 + e^{-2A\omega_c^2t^2}] , &t\ll 1/\omega_c; \\
    \dfrac{1}{2}[1 + (\omega_ct)^{-A}], & 1/\omega_c\ll t\ll 1/T; \\
    \dfrac{1}{2}[1 + e^{-2ATt}], & 1/T\ll t ;\\
\end{array}
\right.\\
\end{split}
\end{equation}

According to the short-time approximation
$Tr_S(\rho^2_S(t))\geqslant 1/2$ at all times. Comparing the
Markovian and the short time approximation one can see in both
cases exponential decay of $Tr_S(\rho^2_S(t))\rightarrow 1/2$ when
$T\rightarrow 0$ and $t\rightarrow\infty$. The difference in
behavior of decay can be explained by the fact that the short-time
approximation is not valid at large times.

I would like to gratefully acknowledge helpful discussions with V.
Privman and D. Averin. This research was supported by the National
Security Agency and Advanced Research and Development Activity
under Army Research Office contract DAAD-19-02-1-0035, and by the
National Science Foundation, grant DMR-0121146.

\end{document}